\documentclass[twocolumn, secnumarabic, amsmath, amssymb, showkeys, nobibnotes, aps, pra, raggedbottom]{revtex4-2}

\usepackage{graphicx}
\usepackage{dcolumn}
\usepackage{bm}
\usepackage{booktabs}

\usepackage{upgreek}
\usepackage{amsmath}
\usepackage{braket}
\usepackage{xcolor}
\usepackage{ulem}
\usepackage{soul}
\usepackage[colorlinks=true, linkcolor=blue, urlcolor=blue]{hyperref}

\begin{document}

\preprint{}

\title{Field demonstration of a self-contained 10 km mobile-to-ground quantum network node via common-aperture co-propagation}

\author{Dongkyu Kim}
\email{dongkyu@add.re.kr}
\author{Dong-Gil Im}
\email{eastgil@add.re.kr}
\author{Kyungdeuk Park}
\author{Jisu Kim}
\author{Jonguk Choi}
\author{Yonggi Jo}
\email{yonggi@add.re.kr}
\author{Yong Sup Ihn}
\email{yong0862@add.re.kr}
\affiliation{Agency for Defense Development, Daejeon 34186, Korea}
\date{\today}

\begin{abstract}
  In this work, we demonstrate a dual-function, common-aperture free-space optical system that simultaneously supports decoy-state BB84 quantum key distribution (QKD) at 803 nm and bidirectional classical laser communication at 640/660 nm over a 10.31 km terrestrial link from a mobile platform to a stationary ground station.
To the best of our knowledge, this is the first field demonstration of a fully integrated common-aperture architecture where a co-propagating classical optical link simultaneously performs full QKD post-processing and encrypted payload transmission over a 10 km atmospheric channel. 
The classical link utilizes Manchester-coded on-off keying (MC-OOK) at 10 MHz, achieving a packet error rate of $\sim$ 0.4 $\%$ (estimated bit error rate (BER) of 3.13 $\times$ 10$^{-5}$) with hardware-based self-synchronization and error-packet retransmission.
Over the 10.31 km link, the QKD system achieved an average quantum bit error rate (QBER) of 2.512 $\%$ and an average secret key rate of 644 bps using rate-adaptive low-density parity-check (LDPC) codes.
Finally, using the distilled keys, real-time AES-256 encrypted image transmission was executed over the co-aligned classical optical channel, proving the feasibility of a completely self-contained, jam resistant mobile quantum optical network.
\end{abstract}


\maketitle

\section{Introduction}

Quantum key distribution (QKD) provides information-theoretic security based on the fundamental laws of quantum mechanics~\cite{Bennett1984IEEE, Ekert1991PRL, Gisin02RMP}.
While fiber-based QKD networks have advanced significantly toward commercialization~\cite{Lucamarini18Nature, Chen21NP}, free-space wireless QKD (W-QKD) remains vital for establishing secure communications with mobile assets—such as unmanned aerial vehicles (UAVs), ground vehicles, maritime vessels, and low Earth orbit (LEO) satellites—where physical optical fibers cannot be deployed~\cite{Pugh17QST, Tian24PRL, Conrad25arXiv, Li25Nature}.

Despite its strategic importance, mobile free-space QKD faces practical operational hurdles. 
First, beam wandering, scintillation, and wavefront distortions caused by dynamic atmospheric turbulence severely degrade spatial coupling efficiency and increase channel loss. 
Second, standard QKD protocols (such as BB84 with decoy states) require a bidirectional classical communication channel for parameter estimation, error reconciliation, and privacy amplification. 
In mobile or military environments, deploying dedicated landlines is impossible, and conventional radio frequency (RF) or commercial cellular links are highly vulnerable to electronic jamming, interception, and relay-node attacks.

Integrating classical free-space optical (FSO) communication directly into the mobile QKD terminal via a shared common-aperture optical transceiver represents an ideal solution. 
By sharing optical tracking and transmission optics, the classical optical link inherently eliminates the vulnerability of RF/cellular channels to electronic jamming and interception, while radically reducing terminal size, weight, power and cost (SWaP-C).
Although the conceptual co-propagation of quantum and classical light has been explored in benchtop or fiber configurations~\cite{Dou23OL, Linale24QCNC, Hua26PR, Dou26PR}, long-range outdoor field demonstration wherein a single shared optical aperture executes both high-precision quantum state transmission and full bidirectional classical optical communication, including complete QKD post-processing distillation and cryptographic data transfer, has not been previously realized.

In this study, we present the end-to-end design, field implementation, and experimental validation of such a system over a 10.31 km mobile-to-station terrestrial link.
We address engineering challenges, including: (1) suppressing chromatic focal shifts to under 0.17 mm across a 150 nm spectral separation via a custom achromatic beam expander; (2) isolating weak single photons from classical co-propagating carrier leakage by $>$ 80 dB; (3) achieving asynchronous, turbulence-resilient clock recovery and automatic frame retransmission using an FPGA-based MC-OOK protocol~\cite{Jing23COP, Zong24Photonics, Ferraro25SPIE}; and (4) executing real-time LDPC reconciliation and privacy amplification entirely over the optical link~\cite{Gallager62IRE, MacKJay96EL, Renner05CC}. 
Finally, we validate the operational integrity of the system through AES-256 tactical image encryption and transmission, demonstrating an autonomous, self-contained quantum-secured mobile communication terminal.

\begin{figure*}[t]
  \centering
  \includegraphics[width=1\textwidth]{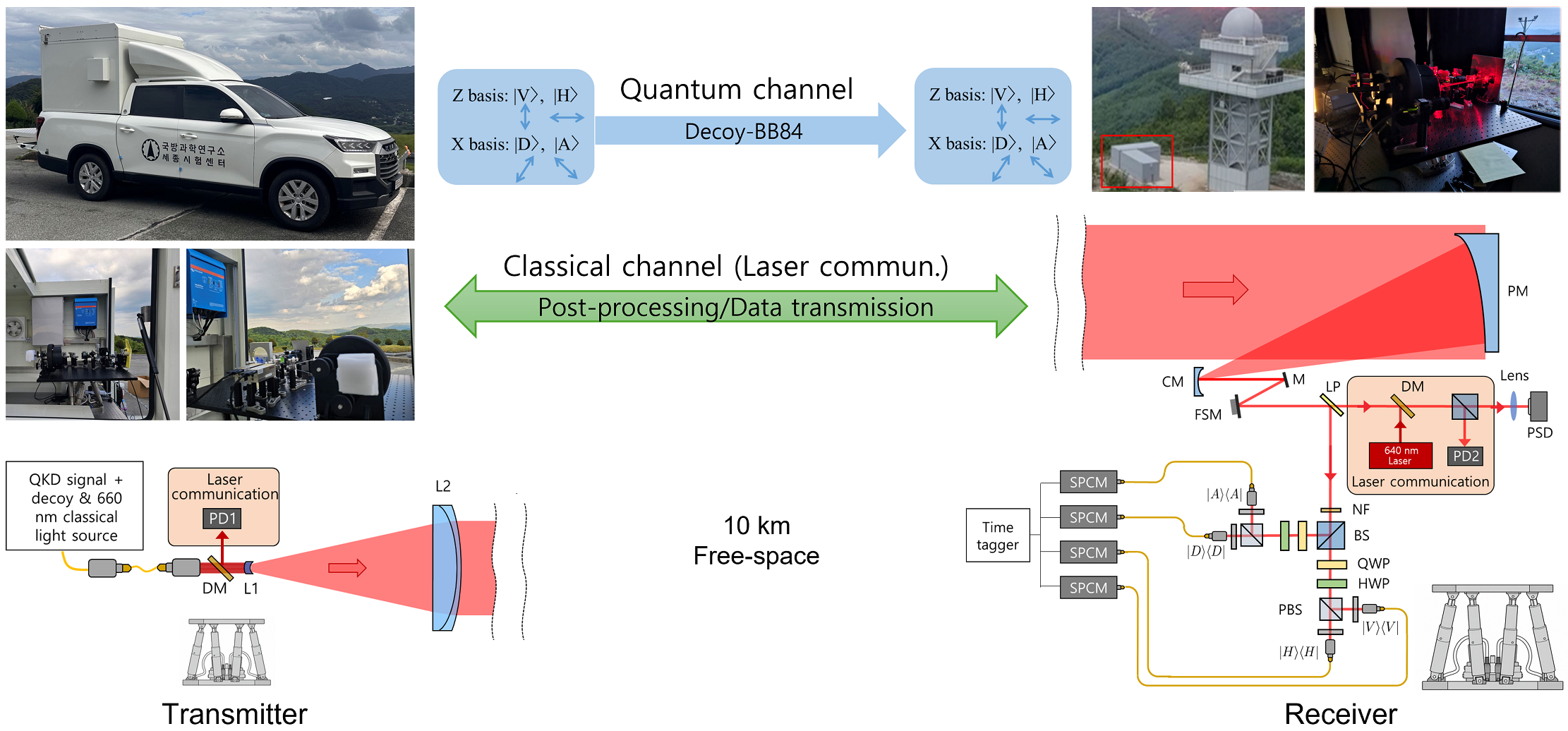}
  \caption{\label{fig1} Schematic and experimental setup of the 10 km mobile-to-station free-space QKD and classical optical communication system. 
Photographs and layouts of the vehicle-mounted transmitter (left) and the tower-based receiver (right). 
The 803 nm decoy-state BB84 quantum channel and 660/640 nm bidirectional classical link share a common aperture for key generation, post-processing, and encrypted data transmission.}
\end{figure*}

\section{Experiments}
The overall system architecture and optical layout of the field-deployable, common-aperture free-space optical (FSO) transceiver are schematically depicted in Fig.~\ref{fig1}. 
The experimental setup establishes a 10.31 km line-of-sight (LOS) propagation channel connecting a mobile vehicle-based transmitter (Alice) and an elevated, fixed optical test tower acting as the receiver terminal (Bob). 
The system simultaneously establishes two co-propagating functional channels: a forward quantum channel transmitting polarization-encoded single-photon states at 803 nm under the decoy-state BB84 protocol, and a full-duplex classical optical communication link utilizing 660 nm (Alice-to-Bob) and 640 nm (Bob-to-Alice) laser carriers.   

\subsection{Optical transmitter on mobile platform (Alice)}
The overall architecture and layout of the mobile transmitter terminal are schematically illustrated in Fig.~\ref{fig1} (left). 
The entire transmitter system—comprising the quantum light sources, classical transceivers, driving electronics, and a 10 kW LiFePO$_{4}$ battery power system—is integrated within a customized aluminum shelter mounted on a four-wheel-drive pickup truck. 
To decouple the optical train from platform vibrations and dynamic field disturbances, the optical assembly is built upon a vibration-isolated optical bench within the mobile enclosure.

The multi-channel quantum transmitter employed in this work follows the experimental setup established in our previous demonstration~\cite{Park26arXiv}.
Specifically, the quantum source utilizes eight independent diode lasers (Toptica iBeam Smart) operating in a pulsed mode at a repetition rate of 100 MHz with a pulse duration of 2.5 ns. 
Under the decoy-state BB84 protocol, four diodes generate signal states with a mean photon number of $\mu=$ 0.8, while the remaining four produce decoy states with $\nu=$ 0.1, combined with vacuum states ($p_{0}$).
To eliminate temporal distinguishability, channel-specific variable optical delay lines compensate for diode-specific internal delays and optical path length mismatches down to $<$ 1 ns.
Four mutually unbiased polarization states across the rectilinear ($\vert H \rangle$, $\vert V \rangle$) and diagonal ($\vert D \rangle$, $\vert A \rangle$) bases are prepared using fiber-based polarization controllers (PCs), polarizing beam splitters (PBSs), and half-wave plates (HWPs) aligned at 22.5$^{\circ}$, yielding an average degree of polarization (DOP) of 98 $\%$.
After passing through a free-space beam combiner, the multiplexed single-mode output is routed through a narrow bandpass tunable filter centered at 803 nm (FWHM $\approx$ 0.5 nm) to eliminate any spectral distinguishability among the eight laser diodes.

Following polarization and spectral preparation, the 803 nm quantum signals are multiplexed with the 660 nm classical optical communication and tracking beam into a common spatial propagation axis using a dichroic mirror (DM). 
To overcome severe geometric beam spreading and diffraction over the 10.31 km atmospheric link, the combined collimated beam ($w_{0}\approx$ 2.6 mm) is launched into a customized two-lens refractive beam expander ($M\approx$ 34$\times$) with an exit pupil diameter exceeding 90 mm.
The expander consists of a negative entrance lens (D-ZF93, $f_{1}=$ -10.44 mm at 660 nm and -10.60 mm at 803 nm) and a large-aperture cemented doublet (pairing D-ZLAF52LA and D-ZF93 glasses, $f_{2}=$ 360.12 mm at 660 nm and 360.29 mm at 803 nm).
By optimizing the relative dispersion characteristics of the two constituent optical glasses, the expander suppresses the inter-wavelength chromatic focal shift to less than 0.17 mm despite the 150 nm spectral separation between the classical and quantum beams. 
This chromatic aberration mitigation ensures that both the 803 nm single photons and the 660 nm classical carrier emerge as near-diffraction-limited, paraxial wavefronts, maximizing link budget stability across the 10.31 km terrestrial free-space channel.

Nevertheless, practical tolerances in the doublet configuration leave an output wavefront curvature radius of $\approx$ 2.21 km immediately after lens 2.
Although this represents a highly collimated state across the 90 mm transmitter aperture, the residual divergence causes the beam diameter to broaden to $\approx$ 450 mm after propagating 10.31 km through the atmosphere, resulting in a geometric expansion that critically influences receiver collection efficiency, as discussed in the following subsection.

\subsection{Stationary optical receiver station (Bob)}
The stationary receiver station was installed at the optical test tower located 10.31 km away from the mobile transmitter, as schematically illustrated in Fig.~\ref{fig1} (right). 
Incoming co-propagating beams traversing the 10.31 km atmospheric link are gathered by a large-aperture off-axis parabolic primary mirror (PM) and redirected via a secondary concave mirror (CM) into the internal optical train. 
The collected light was directed onto a fast-steering mirror (FSM) designed to actively compensate for dynamic beam wander and mechanical vibrations induced by atmospheric turbulence and tower oscillation. 
However, due to significant atmospheric beam broadening over the 10.31 km link, the arriving beam diameter ($\sim$ 450 mm) exceeded the effective receiving aperture of the primary mirror ($\sim$155 mm), causing spatial clipping and thereby limiting the stabilization effectiveness of the FSM system during the trial runs.

Following the FSM, a long-pass (LP) dichroic filter spectrally divides the optical signals. 
The shorter-wavelength classical beams (660 nm forward beacon/data and 640 nm reverse transmitter) are reflected into the classical optical transceiver and position-sensitive detector (PSD) assembly for tracking and data demodulation, whereas the 803 nm quantum signals pass through the filter toward the single-photon detection apparatus.
To prevent high-intensity classical laser scattering and ambient background illumination from saturating the single-photon avalanche diodes (SPADs), a multi-stage optical filtering cascade was implemented. 
The transmitted 803 nm photons pass through a 658/26 nm notch filter (NF) and an 800/12 nm bandpass filter (BF), delivering over 80 dB of out-of-band spectral rejection and reducing optical leakage to well below the single-photon signal threshold.

The filtered single photons enter a passive basis-selection stage governed by a 50:50 non-polarizing beam splitter (BS). 
Polarization analysis along the rectilinear basis ($Z$-basis: ($\vert H \rangle$, $\vert V \rangle$) and diagonal basis ($X$-basis: ($\vert D \rangle$, $\vert A \rangle$) is executed using quarter-wave plates (QWPs), half-wave plates (HWPs), and polarizing beam splitters (PBSs). 
Photons emerging from the PBS output ports undergo an additional stage of 800/12 nm bandpass filtering before being coupled into graded-index multi-mode fibers (MMFs) with a core diameter of 105 $\mu$m.
The fiber outputs are connected to four silicon single-photon counting modules (SPCMs) exhibiting a photon detection efficiency of $\eta\approx$ 60 $\%$ and dark count rates of $\sim$ 1 kcps.
Finally, the detected electrical pulses are routed to a high-precision multi-channel time tagger, which records photon arrival times for coincidence analysis, key sifting, and post-processing.

\subsection{Bidirectional classical optical communication}
Along the classical optical path, the incoming 660 nm forward laser beam is split between a position-sensitive detector (PSD) for active tracking feedback and a high-speed photodetector (PD2) dedicated to classical data demodulation. 
At the same time, a counter-propagating 640 nm laser diode at the receiver terminal is coupled into the identical optical axis to provide reverse data transmission and serve as a tracking reference for Alice’s onboard photodetector (PD1).
Because the mobile transmitter and stationary receiver operate with completely uncoupled, asynchronous master clocks, a Manchester-coded on-off keying (MC-OOK) scheme was implemented using field-programmable gate arrays (FPGAs, National Instruments USB-7856R equipped with a Kintex-7 160T chip). 
The classical laser diodes are driven at an optical modulation rate of 10 MHz, corresponding to an effective data throughput of 5 Mbps. 
In this scheme, the deterministic mid-bit transition inherent to Manchester coding maintains continuous DC baseline balance and provides self-synchronizing clock recovery, effectively preventing low-frequency baseline drifts caused by atmospheric intensity fluctuations.

At the receiver end, the electrical pulses generated from an analog comparator following a high-pass filtered photodetector are oversampled by the FPGA at 40 MHz (4$\times$ oversampling rate relative to the 10 MHz symbol rate).
To reliably extract the data stream, a finite-state-machine (FSM) edge-tracking window was embedded in a Single-Cycle Timed Loop (SCTL). 
A hardware timer tracks intervals between consecutive transitions: edges arriving within a preset count threshold ($\le$ 6 counts) correspond to transition edges between identical bits and are filtered out, while true mid-bit data transitions are successfully decoded.
If no transition occurs within a reset threshold ($\ge$ 13 counts), the tracking timer resets automatically. 
This edge-triggered resynchronization prevents clock drift and phase jitter, ensuring seamless, continuous synchronization even during severe fading and deep signal dropouts caused by atmospheric turbulence.

To interface efficiently with the 64-bit Direct Memory Access (DMA) channel between the FPGA and the host PC while providing resilience against transmission burst errors, a dual-stage packet structure was designed. 
Each transmission set comprises an upper packet and a lower packet (64 bits each, totaling 128 bits per complete set), allocating a 48-bit effective payload alongside a 12-bit synchronization header, a 4-bit control code, a 24-bit sequence identifier, and a 32-bit cyclic redundancy check (CRC-32). 
This framing configuration supports continuous data transfers of up to approximately 100 MB without sequence wrap-around. 
The receiving FPGA performs on-the-fly packet parsing, header identification, and CRC validation at the hardware level, automatically forwarding uncorrupted payloads to the post-processing engine while queuing corrupted sequences for selective automatic repeat request (ARQ) retransmission.

\section{Results and Discussion}
To validate the operational performance and architectural viability of the field-deployable, common-aperture system, experimental evaluations over the 10.31 km terrestrial free-space channel were conducted across four sequential phases: characterization of the atmospheric link budget alongside classical optical communication, execution of decoy-state BB84 QKD with real-time post-processing, verification of end-to-end tactical data encryption using the distilled keys, and an in-depth system analysis of the common-aperture optical transceiver.

\begin{table}[]
\caption{\label{table1} Mesured atmospheric link budget, receiver parameters, and classical communication performance over the 10.31. km channel.}
\renewcommand{\arraystretch}{1.1}
\begin{ruledtabular}
\begin{tabular}{cc}
Parameter & 
\begin{tabular}[c]{@{}c@{}}Experimental value \end{tabular} \\ 
\colrule
Link distance & 10.31 km \\
Weather / Visibility & clear / $>$ 20 km \\
Temperature / Humidity & 20.0 $^{\circ}$C / 79 $\%$ \\
Wind speed / Pressure & $<$ 1 m/s / 1009.8 hPa \\
Atmospheric attenuation & $\sim$ 7.54 $\times$ 10$^{-5}$ m$^{-1}$ \\
Atmospheric transmission & $T_{atm}$ $\sim$ 47 $\%$ \\

\begin{tabular}[c]{@{}c@{}}
Receiver telescope \\ aperture diameter
\end{tabular}
 & $\sim$ 155 mm \\
 
\begin{tabular}[c]{@{}c@{}}
Receiver optics \\ transmission
\end{tabular}
 & $T_{r} \sim 77 \%$ \\

\begin{tabular}[c]{@{}c@{}}
Beam expander \\ geometric coupling
\end{tabular}
 & 15.6 $\pm$ 10.8 $\%$ \\
 
MMF coupling efficiency & $\eta_\text{MMF}$ $\sim$ 50 $\%$ \\

\begin{tabular}[c]{@{}c@{}}
SPCM single-photon \\ detection efficiency
\end{tabular}
 & $\eta_\text{SPCM}$ $\sim$ 60 $\%$ \\
 
SPCM dark count rate & $\sim$ 1,000 cps \\
Background noise & $\sim$ 1,500 cps \\

\begin{tabular}[c]{@{}c@{}}
Intrinsic optical system \\ error (QBER$_\text{IE}$)
\end{tabular}
 & 2 $\%$ \\

\begin{tabular}[c]{@{}c@{}}
Wavelength (Alice $\to$ Bob / \\ Bob $\to$ Alice)
\end{tabular}
 & 660 nm / 640 nm  \\

Modulation scheme &
\begin{tabular}[c]{@{}c@{}}
 Manchester-coded \\ on-off-keying (OOK)
\end{tabular} \\
 
\begin{tabular}[c]{@{}c@{}}
Modulation rate / \\ effective bitrate
\end{tabular}
 & 10 MHz / 5Mbps \\

Effective data throughput & 1.866 Mbps \\
Frame structure & 128 bits total \\
(upper+lower packets) & (48-bit payload) \\
Packet error rate (PER) & $\sim$ 0.4 $\%$ \\
Bit error rate (BER) & 3.13 $\times$ 10$^{-5}$ \\
Transmission time for 10 MB & $\sim$ 44 s \\
\end{tabular}
\end{ruledtabular}
\end{table}

\subsection{Channel budget and classical optical communication performance}
Field trials were executed across a 10.31 km line-of-sight propagation path between the vehicle-mounted mobile transmitter (Alice) deployed at a test site in Sejong, Republic of Korea, and the stationary receiver station (Bob) located at the multi-purpose test tower within the Agency for Defense Development (ADD) in Daejeon.
Meteorological and optical parameters measured during the outdoor experiment are summarized in Table~\ref{table1}.
Experiments were conducted under clear sky conditions with atmospheric visibility exceeding 20 km, an ambient temperature of 20.0 $^{\circ}$C, relative humidity of 79 $\%$, and wind speeds below 1 m/s.
The measured atmospheric attenuation coefficient ($\gamma\approx$ 7.54 $\times$ 10$^{-5}$ m$^{-1}$) corresponded to an atmospheric transmission of $T_{atm}\approx$ 47 $\%$ over the 10.31 km channel. 

Optical characterization of the receiver station revealed an internal optical transmittance of 77 $\%$ and a multi-mode fiber (MMF) core coupling efficiency of approximately 50 $\%$ under stable atmospheric conditions. 
However, from the finite wavefront radius of curvature (2.21 km) at the transmitter expander output, geometric beam spreading expanded the beam profile to $\sim$ 450 mm at the receiver terminal.
Because this spot size considerably exceeded the 155 mm clear aperture of the off-axis parabolic primary mirror, geometric clipping loss occurred at the receiver entrance. 
Over 20 repeated transmission trials, the effective aperture coupling coupling averaged 15.6 $\pm$ 10.8 $\%$ (ranging from 2.2 $\%$ to 39.3 $\%$), with fluctuations primarily driven by uncompensated atmospheric beam wander.
The four silicon single-photon counting modules (SPCMs) exhibited a detection efficiency of $\eta_\text{SPCM}\approx$ 60 $\%$ with dark count rates of about 1 kcps.
By deactivating the visible alignment guide beam (660 nm) during quantum signal transmission, optical leakage into the single-photon detectors was eliminated, leaving ambient background illumination as the noise contributor at $\approx$ 1.5 kcps.

Simultaneously, the bidirectional classical optical communication link was evaluated over the shared optical path to determine its capability to handle autonomous QKD post-processing and cryptographic traffic. 
Modulated via the MC-OOK scheme at an optical symbol rate of 10 MHz (effective bitrate = 5 Mbps), the transceiver sustained a net payload throughput of 1.866 Mbps after accounting for the 128-bit dual-packet framing overhead and inter-packet guard intervals.
For a 10 MB data block (83.9 $\times$ 10$^{6}$ bits, 1,747,627 packets), the forward transmission required approximately 44 seconds without errors.

Across multiple trials, the classical channel demonstrated robust transmission fidelity, recording an average packet error rate (PER) of 0.4 $\%$ to 0.5 $\%$ in the Alice-to-Bob (forward) link and 0.3 $\%$ to 0.4 $\%$ in the Bob-to-Alice (reverse) link, yielding an overall average PER of $\approx$0.4 $\%$.

Assuming uniformly distributed bit errors across the 128-bit frames, this corresponds to an estimated bit error rate (BER) of 3.13 $\times$ 10$^{-5}$. 
Transient packet loss bursts were correlated with rapid beam-wander-induced power fluctuations. 
the embedded selective-repeat automatic repeat request protocol automatically identified and retransmitted corrupted sequence numbers, achieving full error recovery within 1 to 2 seconds. 
These results confirm that the co-propagating classical optical link provides sufficient bandwidth, low latency, and reliability required for real-time basis reconciliation and syndrome exchange without reliance on external RF or commercial telecommunication networks.

\subsection{Quantum key distribution and optical link-based post-processing}
Decoy-state BB84 QKD was performed across the 10.31 km terrestrial link at a laser pulse repetition rate of 100 MHz. 
Using mean photon numbers of $\mu=$ 0.8 for signal states and $\nu=$ 0.1 for decoy states, configured at an emission probability ratio of 2 : 1 : 1 among signal, decoy, and vacuum states, the effective probability ($P_\text{eff}$) of an optical pulse containing at least one photon was 0.3. 
Over 20 transmission trials, each comprising 2.0 $\times$ 10$^{6}$ pulses per 20 ms window, the overall link transmittance excluding receiver aperture clipping was approximately 10.86 $\%$, producing an average raw key rate of 509,093 bps.

The measured quantum bit error rate(QBER) averaged across all basis choices and signal and decoy states was 2.727 $\pm$ 0.551 $\%$.
Under the link parameters, the theoretical error model accounts for detector dark counts ($P_\text{dark}\approx$ 1.0 $\times$ 10$^{-5}$ per gate), ambient background counts ($P_\text{back}\approx$ 1.5 $\times$ 10$^{-5}$ per gate), and intrinsic optical alignment error (QBER$_{IE}\approx$ 0.02) ~\cite{Gisin02RMP}:
\begin{align}
  \text{QBER}_\text{TH}
  &=
  \text{QBER}_\text{IE}+\text{QBER}_\text{dark}+\text{QBER}_\text{back} \\
  &=
\text{QBER}_\text{IE}+\frac{P_\text{dark} n+P_\text{back} n}{2P_\text{eff} T_\text{total} \eta_\text{SPCM}},
  \label{eq1}
\end{align}
where $n$ is the number of detectors. QBER$_\text{TH}$ (2.983 $\%$) shows close agreement with experimentally observed values (2.512 $\%$).
Key distillation metrics and channel parameters are summarized in Table~\ref{table2}.

QKD post-processing was executed in real time over the co-propagating classical optical communication link using a modular post-processing procedure implemented in LabVIEW and optimized C dynamic-link libraries.
The program processes experimental data, estimates the relevant parameters, generates the sifted key, performs error correction, calculates the final secret key length, performs privacy amplification, and outputs the resulting secret key as a binary file.

\begin{table}[t]
\caption{\label{table2} Summary of 10.31 km Decoy-state BB84 QKD and post-processing results.}
\renewcommand{\arraystretch}{1.1}
\begin{ruledtabular}
\begin{tabular}{ccccc}
Parameter & 
\begin{tabular}[c]{@{}c@{}}Experimental value \end{tabular} \\ 
\colrule
Total transmission trials & 106,659,421 \\
Sifted key length & 131,072 bits \\
$Z$-basis signal QBER & 2.547 $\%$ \\
$X$-basis signal QBER & 2.521 $\%$ \\

\begin{tabular}[c]{@{}c@{}}
Single-photon detection \\ lower bound ($s_{Z,1}^{L}$)
\end{tabular}
 & 50,832 bits \\

\begin{tabular}[c]{@{}c@{}}
Single-photon phase error rate \\ lupper bound ($e_{1}^{U}$)
\end{tabular}
 & 8.628 $\%$ \\

\begin{tabular}[c]{@{}c@{}}
Error correction \\ information leakage ($\lambda_\text{EC}$)
\end{tabular}
 & 28,416 bits \\

\begin{tabular}[c]{@{}c@{}}
Total filaure probability ($\epsilon$)
\end{tabular}
 & 10$^{-10}$ \\

\begin{tabular}[c]{@{}c@{}}
Secret key length ($l$)
\end{tabular}
 & 687 bits \\

\begin{tabular}[c]{@{}c@{}}
Quantum transmission duration 
\end{tabular}
 & 1.0666 s \\
 
Secret key rate& 644 bps \\
 
\begin{tabular}[c]{@{}c@{}}
Bidirectional post-processing \\ volume (compressed)
\end{tabular}
 & 148.35 MB \\
 
\begin{tabular}[c]{@{}c@{}}
Optical transfer duration \\ (10 MHz MC-OOK))
\end{tabular}
 & 650 s \\

\begin{tabular}[c]{@{}c@{}}
Post-processing program \\ execution time
\end{tabular}
 & 109 s  \\
 
\begin{tabular}[c]{@{}c@{}}
Total key \\ post-processing time
\end{tabular}
 & 759 s  \\
 
\end{tabular}
\end{ruledtabular}
\end{table}

\begin{figure*}[t]
  \centering
  \includegraphics[width=1\textwidth]{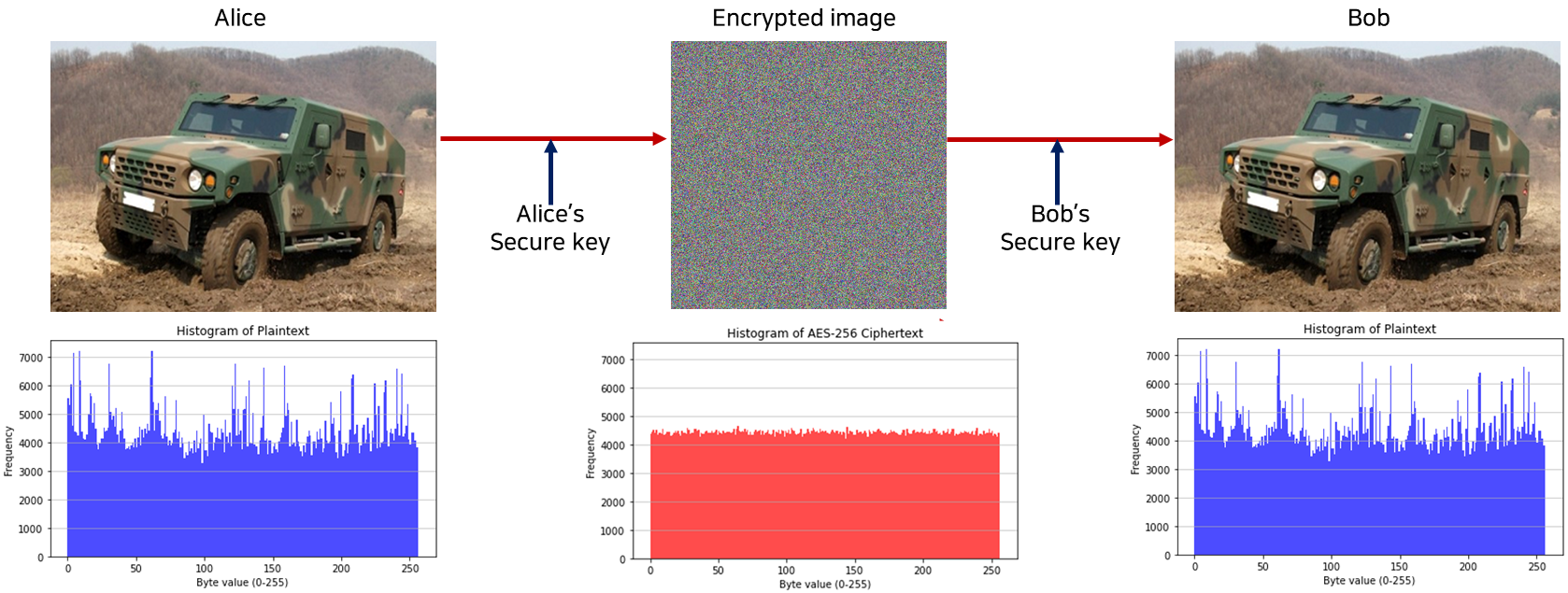}
  \caption{\label{fig2} Demonstration of tactical image encryption, transmission, and decryption using the distributed quantum keys across the 10.31 km optical link.
The panel illustrates the original vehicle image (left), the noise-like encrypted image with header preservation (center), the byte-frequency histograms of the correlated plaintext and uniformly distributed cipertext (bottom), and the decrypted image restored at the receiver (right).}
\end{figure*}

Error reconciliation was conducted using WINNOW \cite{Buttler2003} and low-density parity-check (LDPC) decoding \cite{Gallager1968}. When the QBER is below 0.9\%, WINNOW protocol is performed, otherwise, LDPC decoding is employed. The LDPC matrix is selected according to the measured QBER in 0.25\% intervals below QBER$\leq 3\%$ and 0.5\% interval otherwise.

Privacy amplification was performed using Toeplitz hash matrices, yielding a final verified finite secret key length of 687 bits from a 1.0666 s quantum transmission window, corresponding to an effective secret key rate of 644 bps.

The secret key length is calculated from \cite{Ma05PRA, Zhao06PRL, Gotteman04QIC, Devetak05PRSA}:
\begin{align}
	l=\left\lfloor s_{Z,1}^{L} \left[ 1-H_{2}(e_{1}^{U})\right]-\text{leak}_{\text{EC}}-\Delta(\varepsilon,T) \right\rfloor,
\end{align}
where $s_{Z,1}^{L}$ is the estimated lower bound of the sifted key length corresponding to the events wherer both Alice and Bob used the $Z$-basis and a single photon was emitted and detected, $e_{1}^{U}$ is the estimated upper bound of the $Z$-basis phase error rate for single-photon events, leak$_{\text{EC}}$ is the number of revealed bits during the error correction, and $\Delta(\varepsilon,T)$ is the finite-size correction term with a maximum tolerable failure probability $\varepsilon$ and stopping time $T$. The total failure probability was equally allocated between the correctness and secrecy parameters,  $\varepsilon_{\text{cor}}$ and $\varepsilon_{\text{sec}}$, respectively. The correctness parameter defines the maximum allowable probability that Alice and Bob fail to establish identical keys following reconciliation, while the secrecy parameter quantifies the distinguishability of the generated key from an ideal, uniformly random secret key. We use mixture-martingale bounds for finite-key parameter estimation without assuming independent and identically distributed (i.i.d.) channel statistics \cite{Tamaki2018,Mannalath2026}. Data collection stops when $B=131,072$ sifted signal-intensity $Z$-basis detections have been collected, corresponding to the block length of the LDPC matrix used for error correction. The treatment of this stopping rule is given in the Appendix.

Key distillation metrics and channel parameters are summarized in Table~\ref{table2}. $Z$-basis signal QBER is not estimated during post-processing, since $Z$-basis signals are used to generate the sifted key. This value is reported only for characterization purposes in this paper.

During the classical exchange phase, the raw auxiliary data required for sifting and reconciliation amounted to approximately 1.48 GB per direction, which was compressed to 74.14 MB (Alice-to-Bob) and 74.21 MB (Bob-to-Alice) upon lossless compression, totaling 148.35 MB across 396 bidirectional interactions. 
Transmitted across the 10 MHz classical optical channel, optical transfer required 650 s (including a 2 s retransmission overhead for frame errors), while the computational post-processing took 109 s, resulting in a total execution time of 759 s. 
Although data was processed step-by-step in batches for this feasibility test, running the operations using faster hardware can shorten the optical transfer time to several seconds and support uninterrupted real-time communication.

\subsection{Data encryption and optical transmission demonstration}
To validate the end-to-end operational integrity of the shared quantum cryptographic keys, a real-time data encryption and transmission demonstration was conducted over the 10.31 km classical optical channel, as shown in Fig.~\ref{fig2}.
Using 256 bits of the distilled secret key, a photographic image of a tactical vehicle at the mobile transmitter was encrypted via the Advanced Encryption Standard in Galois/Counter Mode (AES-256).

To verify cryptographic transformation in both spatial and statistical domains, two validation steps were performed. 
First, the image header was preserved while encrypting only the pixel payload, producing the completely randomized, noise-like visual pattern displayed in the center of Fig.~\ref{fig2}. 
Second, to quantitatively evaluate entropy, the entire file, including all headers and structural metadata, was encrypted, and byte-frequency distributions were analyzed across the full 0 to 255 value range. 
As illustrated in the corresponding histograms, while the original plaintext exhibits noticeable peaks representing structural image correlations, the ciphertext histogram displays an ideally flat, uniform distribution across all 256 bins without statistical bias, confirming complete suppression of information leakage.

The ciphertext was subsequently transmitted across the 10.31 km optical channel to the tower receiver. 
Bob successfully decrypted the payload using the matched local key, achieving an image reconstruction. 
This demonstration confirms that the keys distilled via the common-aperture architecture provide information-theoretically robust security capable of protecting real-world tactical payloads in dynamic field scenarios.

\subsection{Integration feasibility and practical scalability}
The co-propagating, common-aperture architecture developed in this work directly addresses the fundamental classical link bottleneck inherent to mobile free-space QKD. 
Conventional free-space systems typically rely on secondary radio-frequency links, directional microwave modems, or commercial cellular networks (such as 4G or 5G) for post-processing communication. 
In contrast, integrating quantum and classical optical channels into a unified physical path offers critical strategic and operational benefits.

First, the classical optical carrier operates with a spatial divergence under 1 mrad, delivering an exceptionally low probability of intercept and detection. 
This makes the classical link completely immune to electronic attacks, radio-frequency tracking, and broadband radio jamming in hostile environments. 
Second, sharing the primary collection mirror, secondary optics, and refractive beam expander between the 803 nm quantum signals and the 640/660 nm classical carriers eliminates duplicate pointing gimbals, independent telescope barrels, and redundant coarse-tracking sensors. 
This substantial reduction in size, weight, power, and cost (SWaP-C) enables practical deployment on compact mobile platforms such as light tactical vehicles. 
Third, co-aligning both beams onto a single optical axis inherently eliminates mechanical parallax and differential thermal misalignment. 
The tracking beam serves as a direct spatial reference for the quantum signal, ensuring that alignment adjustments directly optimize single-photon coupling into the receiving fiber core.

Regarding system throughput and latency, the 650-second transmission duration required for post-processing auxiliary data was bounded strictly by the 10 MHz digital input/output clock limit of the prototyping evaluation FPGA. 
The onboard laser diodes natively support direct modulation bandwidths up to 250 MHz, which would compress classical transfer time to approximately 26 seconds. 
Furthermore, replacing prototyping boards with standard gigabit-capable commercial optical transceivers would reduce auxiliary transmission overhead to under 1 second. 
Integrated with FPGA-accelerated LDPC decoders, this platform is fully capable of sustaining continuous, simultaneous key generation and high-throughput one-time-pad data streaming in dynamic mobile tactical networks.

\section{Conclusion}
In this work, we have designed, implemented, and experimentally demonstrated a field-deployable, common-aperture free-space optical transceiver that simultaneously supports decoy-state BB84 quantum key distribution (QKD) and bidirectional classical laser communication over a 10.31 km terrestrial link. 
Field trials were successfully conducted between a vehicle-mounted mobile transmitter deployed at an outdoor site in Sejong and a stationary receiver station situated at the multi-purpose test tower within the Agency for Defense Development (ADD). To the best of our knowledge, this work represents the first field demonstration of an integrated, common-aperture architecture where a co-propagating classical optical channel executes the entire QKD post-processing workflow, including basis reconciliation, parameter estimation, error correction, and privacy amplification, as well as subsequent encrypted data transmission across an unguided long-range atmospheric channel.

By incorporating a custom achromatic refractive beam expander, the system suppressed the chromatic focal shift between the 803 nm quantum signal and the 660 nm classical carrier to less than 0.17 mm, thereby eliminating mechanical parallax and ensuring near-diffraction-limited collimation. 
Operating over the turbulent 10.31 km outdoor link, the forward quantum channel achieved an average raw key rate of 509.1 kbps with an overall quantum bit error rate (QBER) of 2.512 $\%$.
The classical optical link utilized MC-OOK at 10 MHz to provide hardware-based self-synchronization and selective packet retransmission, maintaining an average packet error rate of approximately 0.4 $\%$ without requiring external RF or commercial network assistance.

Using rate-adaptive LDPC codes and Toeplitz hashing, real-time post-processing distilled a finite secret key rate of 644 bps. 
The practical utility and robust security of the established keys were further validated through the real-time AES-256 encryption, transmission, and flawless recovery of a tactical vehicle image across the co-propagating optical link. 
Statistical analysis confirmed that the transmitted ciphertext achieved an ideally flat byte distribution, verifying maximal entropy and the total elimination of plaintext correlations.

These experimental achievements demonstrate that a fully self-contained, jam-resistant quantum cryptographic network is viable for dynamic mobile platforms operating in contested environments.
Furthermore, because horizontal propagation across 10 km of dense surface atmosphere introduces turbulence and attenuation profiles that meet or exceed those encountered in vertical satellite-to-ground downlinks, the common-aperture methodology verified here provides a robust technological foundation for future multi-domain defense communications, drone-to-ground secure links, and space-based quantum constellations.

\begin{acknowledgments}
  This work was supported by an Agency for Defense Development grant funded by the Defense Acquisition Program Administration.
\end{acknowledgments}

\appendix

\section{Treatment of the Stopping Rule}
In our security analysis, we assume that the receiver, including its click-processing rules, can be described by a basis-independent acceptance filter followed by a squashing map and an ideal BB84 measurement \cite{Gittsovich2014}. The squashing map is a mathematical description, not an additional optical component. This description must reproduce the measurement outcomes and any public information used in the security analysis.

For ideal single-photon BB84 states and the basis-independent receiver model stated above, a phase error in the $Z$-basis key corresponds to an error in a virtual $X$-basis measurement. Independent basis choices therefore give
\begin{equation}
	U^{\mathrm{ph}}_{Z,1,\mu}(n)=\frac{q_Z}{q_X}U^{\mathrm{err}}_{X,1,\mu}(n),
\end{equation}
where $U$ denotes a sum of conditional expectations, $q_Z$ and $q_X$ are the probabilities that both parties choose the corresponding basis, and $n$ is the total number of emitted pulses, including those that do not produce an accepted detection. For the balanced passive receiver, $q_Z/q_X=p_Z/p_X$, where $p_Z$ and $p_X$ are Alice's basis probabilities. We first bound the single-photon $X$-error conditional mean using the decoy data, apply this relation, and then use a further mixture-martingale bound to obtain the upper bound on the $Z$-basis phase-error count. The statistical bounds are included in the failure-probability allocation described below.

Let $N_{Z,\mu}(t)$ be the number of sifted $Z$-basis signal detections obtained from the first $t$ emitted pulses. The pulse count includes all emitted pulses, whether or not they give an accepted detection. Data collection stops at $T=\inf\left\{t:N_{Z,\mu}(t)=B\right\}$, where $B=131,072$.

The estimator uses mixture-martingale bounds on sums of conditional expectations, so the detection and error probabilities may change with the past history \cite{Tamaki2018,Mannalath2026}. To handle the random stopping time, we assign a fixed weight to each possible stopping point:
\begin{align}
	w_n=\frac{B}{n(n+1)}, \quad \sum_{n=B}^{\infty}w_{n}=1.
\end{align}
For a chosen block parameter $\varepsilon$, the estimation and privacy-amplification budgets are 
\begin{align}
	\delta_{n}=\frac{\varepsilon^{2} w_{n}^{2}}{128},\quad \eta_{n}=\frac{\delta_{n}}{26},\quad\alpha_{n}=\frac{\varepsilon w_{n}}{4}.
\end{align}
Here, $\eta_{n}$ is the failure-probability bound for each one-sided bound at the fixed prefix $n$. Since we use 26 such bounds, the union bound limits the probability that any of them fails to $\delta_{n}=26\eta_{n}$. The parameter $\alpha_{n}$ is the trace-distance error budget assigned to privacy amplification for this branch. It sets the shortening term $\Delta_{\text{PA}}(n)=2\log_{2}(1/\alpha_{n})$, i.e., a smaller $\alpha_{n}$ requires a shorter output key.

For each fixed $n$, consider a protocol that processes the first $n$ pulses and aborts unless the first threshold crossing occurs exactly at $n$. Its accepted output is the same as the $T=n$ branch of the stopped protocol. Let $\mathcal{B}_{n}$ be the event that at least one estimation bound fails at this prefix. The union bound gives:
\begin{align}
	\begin{split}
		&\Pr\left[\{T=n\}\cap\mathcal{B}_{n}\right]\le\delta_{n},\\
		&\Pr\left[T<\infty,\mathcal{B}_{T}\right]\le\sum_{n=B}^{\infty}\delta_{n}.
	\end{split}
\end{align}
These probabilities are not conditioned on stopping or accepting. Thus, we do not obtain the stopped bound by simply inserting $T$ into a fixed-sample inequality.

The same branch structure gives the security composition. Let $\rho_{n}$ and $\rho_{n}^{\text{id}}$ be the real and ideal accepted output states for the $T=n$ branch, including the public record and output length. These states keep their branch probability weights. Since $T$ is public, the branches occupy separate classical blocks, and their trace distances add:
\begin{equation}
	D_{\text{sec}}
	=\sum_{n=B}^{\infty}D_{\text{sec},n},\quad D_{\text{sec},n}
	=\frac12\left\|\rho_{n}-\rho_{n}^{\text{id}}\right\|_{1}
\end{equation}
Therefore, if the fixed-prefix security proof gives $D_{\text{sec},n}\le\varepsilon_{\text{sec}}w_{n}$, then $D_{\text{sec}}\le\varepsilon_{\text{sec}}$. The fixed-prefix security analysis includes the public information disclosed during error correction and verification, as described in the maintext \cite{Tupkary2024,Tomamichel2017}.

For the allocation used here, the smoothing and privacy-amplification terms satisfy:
\begin{align}
	2\sqrt{2\delta_{n}}+\alpha_{n}=\frac{\varepsilon w_{n}}{2} \quad \sum_{n=B}^{\infty}\left(2\sqrt{2\delta_{n}}+\alpha_{n}\right)=\frac{\varepsilon}{2}.
\end{align}

The corresponding privacy-amplification shortening term at the observed stopping point is:
\begin{equation}
	\Delta_{\text{PA}}(\varepsilon,T)=2\log_{2}\frac{1}{\alpha_{T}}=2\log_{2}\frac{4}{\varepsilon w_{T}}.
\end{equation}

\end{document}